\documentclass[showpacs,twocolumn,superscriptaddress,notitlepage,preprintnumbers,pra,aps,10pt,floatfix]{revtex4-1}
\usepackage{graphicx} 

\usepackage{amsmath,amssymb,amsthm,mathrsfs,amsfonts,dsfont}
\usepackage{mathdots}
\usepackage{physics}

\usepackage{hyperref}
\hypersetup{colorlinks=true, linkcolor=blue, citecolor=blue, urlcolor=black}
\usepackage{xcolor}

\usepackage{tikz}
\usepackage{physics}
\usepackage[compat=0.6]{yquant}
\usetikzlibrary{quotes}
\usetikzlibrary{fit}
\useyquantlanguage{groups}
\usepackage{multirow}
\usepackage{subcaption}

\newcommand{\ch}[1]{\mathcal{#1}}

\renewcommand{\dag}{^{\dagger}}

\begin{document}

\renewcommand{\dag}{^{\dagger}}

\title{Measurement-Assisted Clifford Synthesis}
\author{Sowmitra Das}
\email{sowmitra.das.sumit@gmail.com}
\affiliation{School of Data and Sciences, BRAC University, Dhaka 1212, Bangladesh}

\begin{abstract}
In this letter, we introduce a method to synthesize an $n$-qubit Clifford unitary $C$ from the stabilizer tableau of its inverse $C\dag$, using ancilla qubits and measurements. The procedure uses ancillary $\ket{+}$ states, controlled-Paulis, $X$-basis measurements and single-qubit Pauli corrections on the data qubits (based on the measurement results). This introduces a new normal form for Clifford synthesis, with the number of two-qubit gates required exactly equal to the weight of the stabilizer tableau, and a depth linear in $n$. 
\end{abstract}

\maketitle

\section{Introduction}
Clifford unitaries are the operators which keep the Pauli operators invariant under the operation of conjugation. The Clifford group is of fundamental importance in the theory of quantum computation, with applications in quantum error correction, quantum compilation and classical simulation of quantum circuits. Although the set of Clifford unitaries are not universal in quantum computation and can be efficiently simulated classically using the Gottesman-Knill theorem~\cite{gottesman1998heisenberg}, if appended with another single gate (most commonly the $T$-gate) it forms the universal Clifford+$T$ gate-set. A key reason for the Clifford group's importance lies in fault-tolerant quantum computation. There exist large classes of stabilizer quantum error-correcting codes in which logical Clifford operators can be implemented transversally, which is the simplest approach to implement gates fault-tolerantly~\cite{kubica2015universal}. Such codes are forerunners in building large-scale quantum computers. As such, compiling unitaries in the Clifford+T gate set is an important primitive in implementing quantum circuits. 

While implementing unitaries, the same functionality an be achieved by different circuits. The problem of generating a sequence of gates from a specification of a given unitary is known as \emph{circuit synthesis}. Which synthesized circuit to use is then determined by available resources and system constraints, such as the number of ancilla qubits, gate-error rates, coherence time, connectivity, and often, just the simplicity of the circuit construction. Given the resource trade-offs inherent in different quantum architectures, it is necessary to have a variety of synthesis techniques available to allow system designers to make appropriate implementation choices. 

The complexity of specifying an arbitrary $n$-qubit unitary grows exponentially with $n$. However, the of an $n$-qubit Clifford unitary can be specified succintly by its stabilizer tableau, which is a table of $O(n^2)$ elements. The problem of synthesizing Clifford unitaries from the tableau specification has been studied in depth and several established techniques exist in the current literature. The most prevalent technique uses appropriate row and column operations on the tableau to reduce it to a canonical form, a number of which have been proposed~\cite{aaronson2004improved, maslov2018shorter, bravyi2021hadamard, bataille2021reduced}. There also exist optimization-based approaches which use SAT solvers~\cite{schneider2023sat, peham2023depth} and template-matching or peephole optimizations~\cite{kliuchnikov2013optimization, bravyi2021clifford}. Recently, graphical methods based on ZX-calculus have been proposed as well~\cite{duncan2020graph, de2025graph}. Optimal depth Clifford circuits have also been found using brute-force search, but the methods only remain feasible for small circuits of up to 26 qubits~\cite{bravyi20226}. Despite these methods, there remains a need for synthesis techniques that offer alternative resource trade-offs, particularly for depth-constrained architectures. 


In this work, we introduce a novel technique for Clifford synthesis,  that trades circuit width for depth, using ancilla qubits and measurements. The circuit is derived from a variation of the recently proposed quantum filter~\cite{das2024purification}, and uses the inverse tableau of the Clifford to be synthesized. This construction gives a new normal form for Clifford circuits consisting layers of ancillary $\ket{+}$ states, controlled-Paulis, $X$-basis measurements and single-qubit Paulis, with a two-qubit gate-cost exactly equal to the number of non-identity elements in the tableau, and a shallow depth of $n+2$. 

In the following sections, we outline the basic algebraic structure of the Clifford group and the stabilizer tableau, outline the circuit for our synthesis technique and the normal form it generates, and provide a proof for the construction. We end with discussions on the advantages of this construction and platforms which might be suitable for implementing it. 

\begin{figure*}[t]
    \resizebox{\textwidth}{!}{
    \begin{tikzpicture}
    \yquantdefinebox{ddots}[inner sep=0pt]{$\ddots$}
    \yquantdefinebox{iddots}[inner sep=0pt]{$\iddots$}
    \yquantdefinebox{vdots}[inner sep=0pt]{$\vdots$}
    \begin{yquant}
    qubit {$a_{1z}: \ket{+}$} q1z; 
    qubit {$a_{1x}: \ket{+}$} q1x;
    qubit {$a_{2z}: \ket{+}$} q2z; 
    qubit {$a_{2x}: \ket{+}$} q2x;
    [register/minimum height=0pt, register/minimum depth=0pt]nobit ellipsis; 
    qubit {$a_{nz}: \ket{+}$} qnz;
    qubit {$a_{nx}: \ket{+}$} qnx;
    qubit {$s:\ket{\psi}$} s1;
    slash {$n$} s1;
    box {$\widetilde{Z_1}$} s1|q1z; 
    box {$\widetilde{X_1}$} s1|q1x; 
    box {$\widetilde{Z_2}$} s1|q2z; 
    box {$\widetilde{X_2}$} s1|q2x; 
    ddots ellipsis; 
    text {$\cdots$} s1;
    box {$\widetilde{Z_n}$} s1|qnz; 
    box {$\widetilde{X_n}$} s1|qnx;
    barrier (q1z, q1x, q2z, q2x, qnz, qnx, s1, ellipsis);
    box {$X_n$} s1|qnx;
    box {$Z_n$} s1|qnz;
    hspace {1pt} ellipsis, s1;
    iddots ellipsis; 
    text {$\cdots$} s1;
    box {$X_2$} s1|q2x;
    box {$Z_2$} s1|q2z;
    box {$X_1$} s1|q1x;
    box {$Z_1$} s1|q1z;
    hspace {1pt} ellipsis, s1;
    vdots ellipsis; 
    measure {$X$} q1z, q1x, q2z, q2x, qnz, qnx; 
    box {$X_1$} s1|q1z;
    box {$Z_1$} s1|q1x;
    box {$X_2$} s1|q2z;
    box {$Z_2$} s1|q2x;
    text {$\cdots$} s1;
    box {$X_n$} s1|qnz;
    box {$Z_n$} s1|qnx;
    discard q1z, q1x, q2z, q2x, qnz, qnx; 
    hspace {5pt} s1;
    output {$C\ket{\psi}$} s1;   
    \end{yquant}
    \end{tikzpicture}
}
\caption{Measurement-Assisted Clifford Synthesis}
\label{fig:big-figure}
\end{figure*}
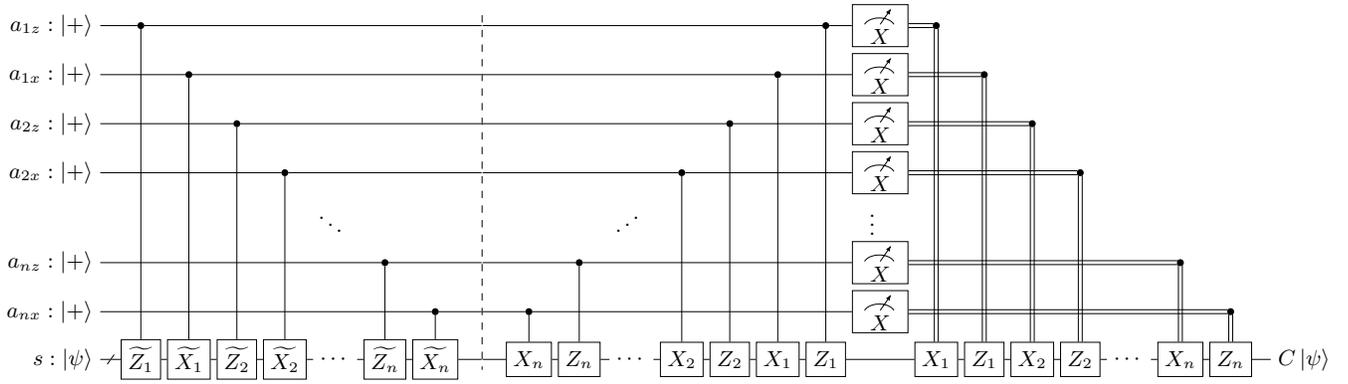

\section{Preliminaries}
In this paper, the following 4 operators will be called the (single-qubit) Pauli operators - 
\begin{align*}
    I = \begin{bmatrix} 1 & 0\\ 0 & 1 \end{bmatrix},
    \quad X = \begin{bmatrix} 0 & 1\\ 1 & 0 \end{bmatrix}, 
    \quad Y = \begin{bmatrix} 0 & -i\\ i & 0 \end{bmatrix}, 
    \quad Z = \begin{bmatrix} 1 & 0\\ 0 & -1 \end{bmatrix}
\end{align*}
The $n$-qubit Pauli group $\mathcal{P}_n$ is formed by $n$-qubit Pauli operators of the form - 
\[i^k P_1\otimes P_2\otimes \cdots \otimes P_n\]
where, $P_i \in \{I, X, Y, Z\}$, $k=0, 1, 2, 3$ and $P_i$ acts on the $i$-th qubit. 

The $n$-qubit Clifford group is the set of $n$-qubit unitaries $C$ such that, for any $P\in \mathcal{P}_n$, $CPC\dag \in \mathcal{P}_n$, i.e, Clifford operators are unitaries which map Pauli operators to Pauli operators (under conjugation). Since the $n$-qubit Paulis form a basis on the space of operators, by linearity, a Clifford unitary $C$ is completely specified by the conjugations $(CPC\dag)_{P\in \mathcal{P}_n}$ (up to a global phase, which is unimportant). And, since the single qubit Paulis $Z_1, X_1, Z_2, X_2, \ldots, Z_n, X_n$ generate $\mathcal{P}_n$, any Clifford $C$ can be further, more succintly, specified by the $2n$ conjugations, 
\[(CZ_iC\dag, CX_iC\dag) \quad i=1, 2, \ldots n\]
The table which lists these $2n$ single-qubit Pauli conjugations is called the stabilizer tableau of $C$. If the stabilizer tableau of a Clifford $C$ is known, then, the inverse tableau listing the conjugations, 
\[(C\dag Z_iC, C\dag X_iC) \quad i=1, 2, \ldots n\]
can also be determined easily~\cite{gidney2021stim}. A brief outline of the inversion procedure is given in \autoref{sec:tableau-inversion}.

In the next section, we show how to construct $C$, given that we know the inverse conjugations $(C\dag Z_i C, C\dag X_i C)_i$. 

\section{Construction}

Let $C$ be an $n$-qubit Clifford unitary. We define, 
\begin{align*}
    \widetilde{Z_i} = C\dag Z_i C\\
    \widetilde{X_i} = C\dag X_i C
\end{align*}
i.e, $\widetilde{Z_i}, \widetilde{X_i}$ are the rows of the stabilizer tableau of the inverse Clifford $C\dag$. Then, the circuit in \autoref{fig:big-figure} implements the Clifford $C$ on the $n$-qubit register $s$, for any arbitrary initial state $\ket{\psi}$ that $s$ is in. The circuit proceeds as follows:  
\begin{enumerate}
    \item First, we introduce $2n$ ancillas in the $\ket{+}$ state. We label the ancillas as $a_{1z}, a_{1x}, a_{2z}, a_{2x}, \ldots,$ $a_{nz}, a_{nx}$.
    \item We apply the Pauli $\widetilde{Z_1} = C\dag Z_1 C$ on the register $s$ (which is in general an $n$-qubit Pauli), controlled on the qubit $a_{1z}$. Then, we apply $\widetilde{X_1}$ on $s$, controlled on $a_{1x}$. We proceed by applying the Paulis $\widetilde{Z_2}, \widetilde{X_2}, \ldots, \widetilde{Z_n}, \widetilde{X_n}$ on $s$ with controls on $a_{2z}, a_{2x}, \ldots, a_{nz}, a_{nx}$ respectively. 
    \item Next we apply the Paulis $X_n, Z_n,$ $X_{n-1}, Z_{n-1}, \ldots, X_2, Z_2, X_1, Z_1$ on $s$ with controls on $a_{nx}, a_{nz}, \ldots, a_{2x}, a_{2z}, a_{1x}, a_{1z}$ respectively. Note that, this layer of controlled gates can be applied in parallel, since the $Z_i, X_i$'s are single qubit Paulis on the register $s$ and the controls are all on seperate qubits. 
    \item We measure all the ancillas in $X$-basis ($\ket{\pm}$ basis). If the the measurement outcome of $a_{iz}$ is (-1), we apply $X_i$ on the $i$-th qubit of $s$, and, if the measurement outcome of $a_{ix}$ is (-1), we apply $Z_i$ on the $i$-th qubit of $s$. This, again, can be done in parallel. 
\end{enumerate}
This completes our implementation of $C$. The circuit consists of the stages $$\rm A-CP_n-CP_1-M_A-P_1$$
which are - (A) initalization of the ancillas (in the $\ket{+}$ state), ($\rm CP_n$) a layer of $n$-qubit  controlled Paulis $\widetilde{Z_i}, \widetilde{X_i}$ determined by the stabilizer tableau of the inverse Clifford $C\dag$, ($\rm CP_1$) a layer of single-qubit controlled Paulis, ($\rm M_A$) measurement of the ancillas and ($\rm P_1$) a layer of single-qubit correction Paulis. This construction thus introduces a new normal form in Clifford synthesis. 

\section{Proof}
\begin{figure*}

\begin{subfigure}{\textwidth}
\begin{tikzpicture}
\begin{yquant}

qubit {$a_{1z}: \ket{+}$} q1z; 
qubit {$a_{1x}: \ket{+}$} q1x;
qubit {$a_{2z}: \ket{+}$} q2z; 
qubit {$a_{2x}: \ket{+}$} q2x;

qubit {$s_1:\hspace{16pt}$} s1;
qubit {$s_2:\hspace{16pt}$} s2;

box {$\widetilde{Z_1}$} (s1,s2)|q1z; 
box {$\widetilde{X_1}$} (s1,s2)|q1x; 

box {$\widetilde{Z_2}$} (s1,s2)|q2z; 
box {$\widetilde{X_2}$} (s1,s2)|q2x; 

box {$X_2$} s2|q2x;
box {$Z_2$} s2|q2z;

box {$X_1$} s1|q1x;
box {$Z_1$} s1|q1z;

measure {$X$} q1z, q1x, q2z, q2x; 

box {$X_1$} s1|q1z;
box {$Z_1$} s1|q1x;

box {$X_2$} s2|q2z;
box {$Z_2$} s2|q2x;

discard q1z, q1x, q2z, q2x;
    
\end{yquant}
\end{tikzpicture}
\caption{}
\end{subfigure}
\hfill
\begin{subfigure}{\textwidth}
\begin{tikzpicture}
\begin{yquant}

qubit {} q1z; 
qubit {} q1x;
qubit {} q2z; 
qubit {} q2x;

discard q1z, q1x, q2z, q2x; 

qubit {$s_1:$} s1;
qubit {$s_2:$} s2;

box {$C$} (s1 ,s2);

barrier (q1z, q1x, q2z, q2x, s1, s2);

init {$a_{1z}: \ket{+}$} q1z; 
init {$a_{1x}: \ket{+}$} q1x;
init {$a_{2z}: \ket{+}$} q2z; 
init {$a_{2x}: \ket{+}$} q2x;

box {$Z_1$} s1|q1z; 
box {$X_1$} s1|q1x; 

box {$Z_2$} s2|q2z; 
box {$X_2$} s2|q2x; 

hspace {5pt} s1, s2;
[fill=red!20, "$\ch{E}$" below]box {$C\dag$} (s1, s2);
hspace {5pt} s1, s2;

box {$X_2$} s2|q2x;
box {$Z_2$} s2|q2z;

box {$X_1$} s1|q1x;
box {$Z_1$} s1|q1z;

measure {$X$} q1z, q1x, q2z, q2x; 

box {$X_1$} s1|q1z;
box {$Z_1$} s1|q1x;

box {$X_2$} s2|q2z;
box {$Z_2$} s2|q2x;

discard q1z, q1x, q2z, q2x;
    
\end{yquant}
\end{tikzpicture}
\caption{}
\end{subfigure}
\caption{(a) Synthesis circuit for a 2-qubit Clifford. (b) Equivalent circuit.}
\label{fig:equiv-circuit}
\end{figure*}
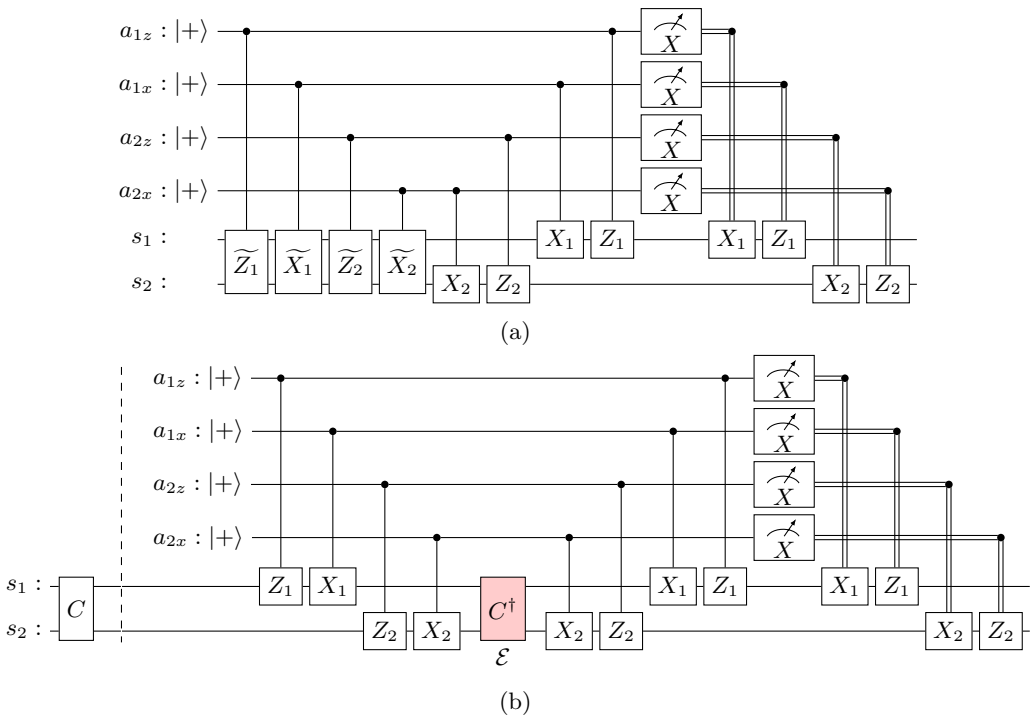

For ease of drawing figures, we show the proof for the 2-qubit case, which is easily generalized to $n$ qubits. For two qubits, the circuit in \autoref{fig:big-figure} looks as the circuit in \autoref{fig:equiv-circuit}(a)  below, with the data qubits shown seperately.


























For each $\widetilde{X_i} = C\dag X_i C$ (and similarly for $\widetilde{Z_i}$), we have the following reduction, 

\begin{center}
\begin{tikzpicture}
\begin{yquantgroup}
\yquantset{operator/separation=2mm}
    \registers{
    qubit {} q1;
    qubit {} s1;
    qubit {} s2;
    }
    \circuit{
    box {$C\dag X_2C$} (s1, s2)|q1; 
    }
    \equals
    \circuit{
    box {$C$} (s1, s2); 
    box {$X_2$} s2 | q1;
    box {$C\dag$} (s1, s2); 
    }
\end{yquantgroup}
\end{tikzpicture}
\end{center}

Using this identity, the circuit \autoref{fig:equiv-circuit} (a)  reduces to the circuit in \autoref{fig:equiv-circuit} (b).

















    

It was shown in the quantum filter construction in \cite{das2024purification} that, for any channel $\mathcal{E}$ (which in this circuit is the unitary $C\dag$), the circuit (to the right of the dashed line) above reduces to the identity channel (a proof of this is given in \autoref{app:quantum-filter}). Thus, the resulting circuit is equivalent to just applying $C$ on the register $s$, and our proof is complete. 

\section{Outlook}
In this work, we have outlined a method to construct a Clifford unitary using ancilla qubits and auxiliary measurements. The construction is conceptually simple, and does not require any modifications on the stabilizer tableau except computing its inverse. The construction requires $2n$ ancilla qubits, but, has a shallow depth of $(n+2)$. Interestingly, no 2-qubit gates are applied between the qubits internal to the ancilla and data registers - all the 2-qubit gates have a control on the ancilla register and target on the data register. As such, it does not require any intra-connections within the ancilla and data registers. However, it does require all-to-all interconnections between the two registers, and, would thus be more suitable for trapped-ion systems. 

\widetext
\appendix

\section{Stabilizer Tableau Inversion}
\label{sec:tableau-inversion}
We briefly outline how one can find the stabilizer tableau of the inverse of a Clifford, $C\dag$, given that the tableau of $C$ is known. For more details, one can look into standard references such as~\cite{aaronson2004improved, gidney2021stim}. 

Any $n$-qubit Pauli-string $P_1\otimes P_2\otimes \cdots \otimes P_n$ may be represented as a bipartite vector of the form - 
\[P_1\otimes P_2\otimes \cdots \otimes P_n \longrightarrow (z_1\; z_2 \cdots z_n \mid x_1\; x_2 \cdots x_n)\]
where, $z_ix_i = (00, 01, 10, 11)$ if $P_i = (I, X, Z, Y)$ respectively. 

Conjugating a Pauli-string $P$ by a Clifford $C$ results in another Pauli-string $Q$, up to a phase of $\pm 1$, i.e, 
\[CPC\dag = s\cdot Q, \quad s=\pm 1\]
Hence, the $2n$ Pauli-conjugations $(CZ_iC\dag, CX_iC\dag)$ for $i=1, 2, \ldots, n$ may each be represented by a vector of length $2n+1$ appending the sign-bit, 
\[CZ_iC\dag \longrightarrow (s_{iz} \mid z_{iz, 1}\; z_{iz, 2} \ldots z_{iz, n}\mid x_{iz, 1} \; x_{iz, 2}\cdots x_{iz, n})\]
and similarly for $CX_iC\dag$. Listing these $2n$ vectors one below the other, gives us a matrix of size $2n\times (2n+1)$ which represents the stabilizer tableau of $C$. The binary part of this matrix (excluding the sign-bits) is a symplectic matrix of the block-form 
\begin{equation}
    \left[\begin{array}{c|c}
         \mathbf{z}_{_Z}& \mathbf{x}_{_Z} \\[2pt]
         \hline
         \mathbf{z}_{_X}& \mathbf{x}_{_X}
    \end{array}\right]
\end{equation}
A binary symplectic matrix as this one is easily inverted, with the inverse given by - 
\begin{equation}
\renewcommand{\arraystretch}{1.8}
    \left[\begin{array}{c|c}
         \mathbf{x}_{_X}^{\sf T}& \mathbf{x}_{_Z}^{\sf T}\\
         \hline
         \mathbf{z}_{_X}^{\sf T}& \mathbf{z}_{_Z}^{\sf T}
    \end{array}\right]
\end{equation}
Thus, the binary part of the inverse tableau, i.e, the Pauli part of the inverse conjugations ($C\dag Z_iC, C\dag X_i C$) is simply determined by these matrix transpositions. All that is left is to determine the sign of the inverse conjugations. Suppose, the Pauli part of the conjugation $C\dag Z_iC$ is $Q_{iz}$, and the corresponding sign is $\tilde{s}_{iz}$, i.e, 
\begin{equation}
    C\dag Z_iC = \tilde{s}_{iz}Q_{iz}
\end{equation}
Then, $\tilde{s}_{iz}$ is given by, 
\begin{equation}
    \tilde{s}_{iz}I = Z_i\cdot CQ_{iz}C\dag
\end{equation}
Since the entire tableau of $C$ is known, we can compute the Pauli $CQ_{iz}C\dag$ using the original tableau. Multiplying this by $Z_i$ gives us $\tilde{s}_{iz}$. Interestingly, most of the effort in computing the inverse tableau is spent in computing the signs. 

\section{Quantum Filter}
\label{app:quantum-filter}

Here, we provide a simple proof that the quantum filter circuit in \autoref{fig:equiv-circuit} reduces to the identity channel for unitary operator $C\dag$. We will show this by proving the equivalence for Pauli operators, and the final result follows from the fact that Pauli operators form a basis on the space of operators. We shall make another simplification of proving the equivalence for a 1-qubit quantum filter (with 2 ancilla qubits). The proof is easily generalized to multiple qubits. For further details on the construction of the quantum filter, the reader may refer to~\cite{das2024purification}. 

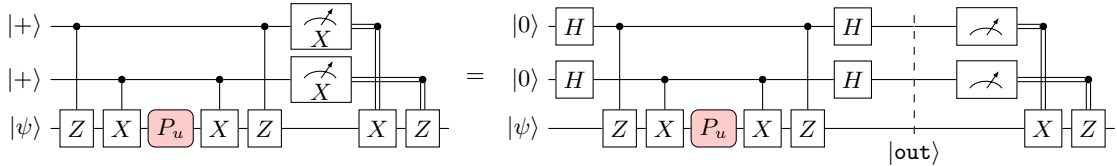
\begin{figure}[!h]
    \begin{center}
    \begin{tikzpicture}
    \begin{yquantgroup}
    \registers{
    qubit {} q1;
    qubit {} q2;
    qubit {} r;
    }
    \circuit{
    init {$\ket{+}$} q1;
    init {$\ket{+}$} q2;
    init {$\ket{\psi}$} r;
    box {$Z$} r|q1; 
    box {$X$} r|q2; 
    [shape=yquant-rectangle, rounded corners=.3em, fill=red!20, x radius= 0.3cm]box {$P_u$} r; 
    
    box {$X$} r|q2 ;  
    box {$Z$} r|q1;

    measure {$X$} q1, q2; 
    box {$X$} r|q1;
    box {$Z$} r|q2;
    discard q1, q2;
    }
    \equals
    \circuit{
    init {$\ket{0}$} q1;
    init {$\ket{0}$} q2;
    init {$\ket{\psi}$} r;
    h q1, q2;
    box {$Z$} r|q1; 
    box {$X$} r|q2; 
    [shape=yquant-rectangle, rounded corners=.3em, fill=red!20, x radius= 0.3cm]box {$P_u$} r; 
    
    box {$X$} r|q2 ;  
    box {$Z$} r|q1;
    h q1, q2;

    ["$\ket{\tt out}$" below] barrier (q1, q2, r);
    
    measure q1, q2; 
    box {$X$} r|q1;
    box {$Z$} r|q2;
    discard q1, q2;
    }
    \end{yquantgroup}
    \end{tikzpicture}
    \end{center}
    \caption{Single-qubit quantum filter}
    \label{fig:single-qubit-channel-correction}
\end{figure}

Consider the circuit in \autoref{fig:single-qubit-channel-correction}, where a Pauli unitary acts in the middle of the filter. For convenience, we define $(P_{00}, P_{01}, P_{10}, P_{11}) = (I, Z, X, Y)$. One can easily check that, for $P_u = I, Z, X, Y$ respectively, the state labelled $\ket{\tt out}$ in the circuit on the right is - 
\begin{align*}
    P_u = (I, Z, X, Y) \longrightarrow \ket{\tt out} &= (\ket{00}\otimes \ket{\psi}, \ket{01}\otimes \ket{\psi}, \ket{10}\otimes \ket{\psi}, \ket{11}\otimes \ket{\psi})\\
    & = \ket{u}\otimes P_u\ket{\psi}
\end{align*}
where, $u=00, 01, 10, 11$ as above. Thus the filter maps an input state $\ket{\psi}$ to the output state, 
\begin{equation}\label{eq:1-qubit-full-filter}
    \ket{\psi} \to \ket{\tt out} = \ket{u} \otimes P_u\ket{\psi}
\end{equation}
Now, assume that $P_u$ is replaced by a channel $\ch{E}$ which is described by Kraus elements $\{E_k\}$, i.e, $\ch{E}(\rho) = \sum_k E_k\rho E_k$. Since the Pauli operators form a basis on the space of operators, we can write the $E_k$'s as 
\begin{equation}
    E_k = \sum_{u \in (\mathbb{Z}_2)^2} \alpha_{k, u} P_u 
\end{equation}
where the sum is over all the (single-qubit) Paulis. Hence, if $P_u$ is replaced by any of the $E_k$'s in \autoref{fig:single-qubit-channel-correction}, by linearity, the (unnormalized) output-state will be 
\begin{equation}
    \ket{\psi} \xrightarrow{E_k} \ket{{\tt out}_k} = \sum_{u \in (\mathbb{Z}_2)^2} \alpha_{k, u} \ket{u}\otimes P_u \ket{\psi}
\end{equation}

Therefore, the normalized output state for the channel $\ch{E}$ plugged into the filter is 
\begin{align}
    \rho_{\tt out} &= \sum_k \dyad{{\tt out}_k}\\
    &= \sum_k \left(\sum_{u} \alpha_{k, u} \ket{u}\otimes P_u \ket{\psi}\right) \left(\sum_{v} \alpha_{k, v}^* \bra{v}\otimes P_v \bra{\psi}\right) \\
    &= \sum_k \sum_{u, v} \alpha_{k, u}\alpha_{k, v}^* \ketbra{u}{v} \otimes P_u (\dyad{\psi}) P_v
\end{align}

Now, in the measurement phase, the ancillas are measured in the computational basis and a corresponding Pauli is applied to the data qubit (similar to a recovery operation). According the circuit in the figure, if the measurement result on the two ancillas is $\ket{u_0}\otimes \ket{u_1} = \ket{u}$, then the corresponding correction Pauli is $X^{u_0}Z^{u_1} = P_u$ (up to a global phase, which is unimportant). If the measurement projectors are $\mathit{\Pi}_u = \dyad{u}$, the recovery channel is given by 
\begin{align}
    \ch{C}(\rho) = \sum_u (\mathit{\Pi}_u \otimes P_u) \rho (\mathit{\Pi}_u \otimes P_u)^{\dagger}
\end{align}
Therefore, the post-recovery state is 
\begin{align}
    \ch{C}(\rho_{\tt out}) &= \sum_k \sum_{u, v, w} \alpha_{k, u}\alpha_{k, v}^* (\mathit{\Pi}_w\ketbra{u}{v}\mathit{\Pi}_w) \otimes P_w(P_u \dyad{\psi}P_v)P_w\\
    &= \left(\sum_k \sum_u \abs{\alpha_{k, u}}^2 \dyad{u}\right) \otimes \dyad{\psi}
\end{align}
Hence, the system qubit is returned to its original state. Thus, the entire filter circuit acts like an identity channel on the data qubit $\ket{\psi}$ for any channel $\ch{E}$ and the proof is complete. The proof is easily extended for mixed input states. And for the general case of multiple qubits, one can just use a filter for each qubit separately, and repeat the procedure given above. 

\bibliographystyle{unsrt}
\bibliography{reference}

\end{document}